\documentclass[a4paper,12pt]{article}
\usepackage[english]{babel}
\usepackage{graphicx,color}
\usepackage{ifpdf}
\ifpdf
  \usepackage[pdftex]{hyperref}
\else
  \usepackage[hypertex]{hyperref}
\fi
\hypersetup{bookmarks=true, unicode=true, colorlinks=true, linkcolor=black, citecolor=black, filecolor=black, urlcolor=black}
\ifpdf
  \DeclareGraphicsExtensions{.jpg}
\else
  \DeclareGraphicsExtensions{.eps}
\fi
\usepackage{amsmath,amssymb,amsfonts,bm,euscript,amscd,ae,amsthm,exscale,amsopn,amstext}
\usepackage{ifthen}
\usepackage[T1]{fontenc}
\usepackage[cp1251]{inputenc}
\usepackage[active]{srcltx}

\hypersetup{colorlinks,%
linkcolor=blue,%
urlcolor=blue,%
citecolor=red,%
filecolor=magenta}

\title{Preliminaries in Many-Particle Quantum Gravity: Einstein--Friedmann Spacetime}

\author{Lukasz Andrzej Glinka\footnote{E-mail address: laglinka@gmail.com}}
\date{\today}

\begin{document}

\maketitle
\begin{abstract}
Preliminaries for Many-Particle approach to quantization of Einstein--Hilbert theory of gravitation are presented in this paper. Einstein--Friedmann Spacetime is detailed discussed from this point of view. Von Neumann--Araki--Woods second quantization in Fock space of canonical Hamilton field equations is presented. In result Spacetime as system with Bose--Einstein statistics is described and thermodynamics of the system is computed.
\end{abstract}

\section{Introduction}
In this paper introductory results of Fock space formulation of Quantum Gravity are discussed by Einstein--Friedmann Spacetime primer. Firstly classical Dirac Hamiltonian approach to General Relativity for the Spacetime is briefly presented, and Wheeler--DeWitt evolution as primary quantization of constraints is formulated. Next evolution is separated on canonical Hamilton field equations system, and consistent von Neumann--Araki--Woods quantization in Fock space is realized. For diagonalization Bogoliubov--Heisenberg basis is used, and internal logarithmic conformal field theory structure in this basis is proved. By treatment of the Spacetime as Bose--Einstein system thermodynamics of the system is computed.
\section{Classical approach}
\subsection{Dirac constraints}
Flat, homogenous and isotropic Spacetime investigated by Einstein \cite{e} and Friedmann \cite{f} is given by interval
\begin{equation}\label{ef}
ds^{2}=a^{2}(\eta)\left[(d\eta)^2-(dx^i)^2\right],~~d\eta=N_d(x^0)dx^0,
\end{equation}
here $a(\eta)$ is Friedmann conformal scale factor, $\eta$ is conformal time in Dirac sense \cite{d}, and $N_d$ is lapse function. Hilbert action \cite{h} for (\ref{ef}) can be written as
\begin{equation}\label{hil}
\mathcal{A}[a]=\int{dx^0}\left\{p_a\dfrac{da}{dx^0}-N_d\left[-\dfrac{p_{a}^{2}}{4V_0^2}+\rho(a)\right]V_0\right\},
\end{equation}
where $p_a$ is nontrivial canonical conjugate momentum, $\rho(a)$ is energy density
\begin{eqnarray}
p_{a}=-\dfrac{2V_0}{N_d}\dfrac{da}{dx^0},~~\rho(a)=\dfrac{a^4}{V_0}\int{d^{3}x}~\mathcal{H}_{M},
\end{eqnarray}
and $V_0=\int{d^{3}}x<\infty$ is finite space volume. Variational principle with respect to $N_d$ applied to action (\ref{hil}) gives constraints
\begin{equation}\label{c}
\dfrac{\delta\mathcal{A}[a]}{\delta N_d}=0\Rightarrow-\dfrac{p_{a}^{2}}{4V_0^2}+\rho(a)=0,
\end{equation}
with solution given by Hubble law
\begin{equation}
\int_{a_I}^{a}\dfrac{da}{\sqrt{\rho(a)}}=\pm|\eta-\eta_I|,
\end{equation}
where index $I$ means initial data.
\subsection{Hamilton field equations}
One can say that Dirac constraints (\ref{c}) have a form
\begin{equation}\label{4.1}
p_{a}^{2}-{\omega}^{2}(a)=0,~~\omega(a)\equiv\pm2V_0\sqrt{\rho(a)},
\end{equation}
and that direct primary quantization
\begin{equation}\label{4.2}
i\left[\hat{\mathrm{p}}_{a},a\right]=1,~~\hat{\mathrm{p}}_{a}\equiv-i\dfrac{\partial}{\partial{a}},
\end{equation}
gives Wheeler--DeWitt evolution \cite{w1,w2}
\begin{equation}\label{wdw}
\left(\dfrac{\partial^2}{\partial{a^2}}+{\omega}^{2}(a)\right)\Psi(a)=0.
\end{equation}
Equation (\ref{wdw}) can be represented as canonical Hamilton field equations system
\begin{equation}\label{ham}
\left\{\begin{array}{ccc}\dfrac{\partial}{\partial\Pi_{\Psi}}\mathrm{H}(\Pi_{\Psi},\Psi)\!\!\!\!\!&=&\!\!\!\!\!\partial_{a}\Psi,\vspace*{5pt}\\
-\dfrac{\partial}{\partial\Psi}\mathrm{H}(\Pi_{\Psi},\Psi)\!\!\!\!&=&\!\partial_{a}\Pi_{\Psi},
\end{array}\right.
\end{equation}
where considered Hamiltonian $\mathrm{H}(\Pi_{\Psi},\Psi)$ has a form
\begin{equation}\label{H}
\mathrm{H}(\Pi_{\Psi},\Psi)=\dfrac{1}{2}\left(\Pi^{2}_{\Psi}+{\omega}^{2}(a)\Psi^{2}\right).
\end{equation}
In further text we will use compact form of (\ref{ham})
\begin{equation}\label{ham1}
\dfrac{\partial}{\partial{a}}\left[\begin{array}{c}\Psi\\
\Pi_\Psi\end{array}\right]=\left[\begin{array}{cc}
0&1\\
-{\omega}^{2}(a)&0\end{array}\right]\left[\begin{array}{c}\Psi\\
\Pi_\Psi\end{array}\right].
\end{equation}

\section{Quantization}
We focus attention on Hamilton field equations (\ref{ham1}).
\subsection{Von Neumann--Araki--Woods quantization}
Essence of quantization constitutes general transition between classical fields and field operators
\begin{equation}\label{tra}
\left[\begin{array}{c}\Psi(a)\\\Pi_\Psi(a)\end{array}\right]\longrightarrow\left[\begin{array}{c}\mathbf{\Psi}[a(\eta)]\\\mathbf{\Pi_\Psi}[a(\eta)]\end{array}\right],
\end{equation}
with Canonical Commutation Relations (CCRs)
\begin{equation}\label{ccr}
\left[\mathbf{\Pi_{\Psi}}[a(\eta)],\mathbf{\Psi}[a(\eta')]\right]=-i\delta\left(a(\eta)-a(\eta')\right).
\end{equation}
Problem of CCRs representations is not new, and was investigated in \cite{ccr1} and \cite{ccr2}. In \cite{g} was proposed von Neumann--Araki--Woods quantization in Fock space $\left(\mathcal{G},\mathcal{G}^{\dagger}\right)$ of annihilation and creation operators
\begin{equation}\label{2nd}
\left[\begin{array}{c}\mathbf{\Psi}\\\mathbf{\Pi_\Psi}\end{array}\right]
=\left[\begin{array}{cc}\dfrac{1}{\sqrt{2{\omega}(a)}}&\dfrac{1}{\sqrt{2{\omega}(a)}}\\
-i\sqrt{\dfrac{{\omega}(a)}{2}}&i\sqrt{\dfrac{{\omega}(a)}{2}}\end{array}\right]
\left[\begin{array}{c}\mathcal{G}[a(\eta)]\\ \mathcal{G}^{\dagger}[a(\eta)]\end{array}\right],
\end{equation}
which is consistent with (\ref{ccr}) by CCRs
\begin{eqnarray}
\left[\mathcal{G}[a(\eta)],\mathcal{G}^{\dagger}[a(\eta')]\right]&=&\delta(a(\eta)-a(\eta')),\label{ccrs1}\\
\left[\mathcal{G}[a(\eta)],\mathcal{G}[a(\eta')]\right]&=&0\label{ccrs2}.
\end{eqnarray}
Quantization (\ref{2nd}) applied to equations (\ref{ham1}) gives
\begin{equation}\label{4.11a}
\dfrac{\partial}{\partial{a}}\left[\begin{array}{c}\mathbf{\Psi}\\
\mathbf{\Pi_\Psi}\end{array}\right]\!=\!\left[\begin{array}{cc}
0&1\\
-{\omega}^2(a)&0\end{array}\right]\left[\begin{array}{c}\mathbf{\Psi}\\
\mathbf{\Pi_\Psi}\end{array}\right],
\end{equation}
or in terms of Fock space $\left(\mathcal{G},\mathcal{G}^{\dagger}\right)$
\begin{equation}\label{fock}
\dfrac{\partial}{\partial{a}}\left[\begin{array}{c}\mathcal{G}[a(\eta)]\\
\mathcal{G}^{\dagger}[a(\eta)]\end{array}\right]\!=\!\left[\begin{array}{cc}
i{\omega}(a) & \Delta\\
\Delta&
-i{\omega}(a)\end{array}\right]\left[\begin{array}{c}\mathcal{G}[a(\eta)]\\
\mathcal{G}^{\dagger}[a(\eta)]\end{array}\right],
\end{equation}
where $\Delta=\partial_{a}\ln\left|\dfrac{\omega(a)}{\omega(a_I)}\right|$, and $\omega(a_I)=\pm4V_0\sqrt{\rho(a_I)}$ is initial value of $\omega(a)$.
\subsection{Bogoliubov--Heisenberg operator basis}
Evolution (\ref{fock}) should be diagonalize in order to build correct quantum theory. By character of evolution (\ref{wdw}) we apply Bose--Einstein type Bogoliubov automorphism
\begin{eqnarray}
\left[\begin{array}{cc}\mathcal{W}[a(\eta)]\\
\mathcal{W}^{\dagger}[a(\eta)]\end{array}\right]&=&\left[\begin{array}{cc}u(a)&v(a)\\
v^{\ast}(a)&u^{\ast}(a)\end{array}\right]\left[\begin{array}{cc}\mathcal{G}[a(\eta)]\\
\mathcal{G}^{\dagger}[a(\eta)]\end{array}\right],\label{bog1}\\
&&|u(a)|^{2}-|v(a)|^{2}=1,\label{hip}
\end{eqnarray}
which lies in accordance with CCRs (\ref{ccrs1}) and (\ref{ccrs2})
\begin{eqnarray}
\left[\mathcal{W}[a(\eta)],\mathcal{W}^{\dagger}[a(\eta')]\right]&=&\delta(a(\eta)-a(\eta')),\label{ccr1}\\
\left[\mathcal{W}[a(\eta)],\mathcal{W}[a(\eta')]\right]&=&0\label{ccr2},
\end{eqnarray}
and we reduce evolution (\ref{fock}) to Heisenberg one
\begin{equation}\label{4.15}
\dfrac{\partial}{\partial{a}}\left[\begin{array}{c}\mathcal{W}[a(\eta)]\\
\mathcal{W}^{\dagger}[a(\eta)]\end{array}\right]=\left[\begin{array}{cc}
i\lambda& 0 \\ 0 &
-i\lambda\end{array}\right]\left[\begin{array}{c}\mathcal{W}[a(\eta)]\\
\mathcal{W}^{\dagger}[a(\eta)]\end{array}\right].
\end{equation}
In result we obtain $\lambda=0$, $\mathcal{W}\equiv\mathrm{w}=\mathit{constant}$, and vacuum state $|0\rangle$ defined by relation
\mbox{$\mathrm{w}|0\rangle=0$}. Evolution (\ref{fock}) is equivalent to equations on functions $u$ and $v$
\begin{equation}\label{4.25}
\dfrac{\partial}{\partial{a}}\left[\begin{array}{c}v(a)\\u(a)\end{array}\right]=\left[\begin{array}{cc}i{\omega}(a)&\Delta\\
\Delta&-i{\omega}(a)\end{array}\right]\left[\begin{array}{c}v(a)\\u(a)\end{array}\right],
\end{equation}
that can be solved in hyperbolic parametrization
\begin{eqnarray}
v(a)=e^{i\theta(a)}\sinh \phi(a),~~u(a)=e^{i\theta(a)}\cosh \phi(a).
\end{eqnarray}
Here functions--parameters
\begin{eqnarray}
\theta(a)=\int_{a_I}^{a}p_ada,~\phi(a)=\ln{\left|\dfrac{\omega(a_I)}{\omega(a)}\right|}.
\end{eqnarray}
are classical phase integral and logarithmic conformal field. In this manner we have proved that theory builded von Neumann--Araki--Woods quantization in form (\ref{2nd}) has internal logarithmic conformal field theory structure.
\section{Physical implications}
Now some physical results will present.
\subsection{Density functional}
We take density functional operator in standard form of occupation number consistent with quantization (\ref{2nd})
\begin{equation}
\varrho=\mathcal{G}^{\dagger}\mathcal{G},
\end{equation}
which in Bogoliubov--Heisenberg basis has a following matrix representation
\begin{equation}
\varrho=\left[\begin{array}{cc}|u|^2&-uv\\-u^{\ast}v^{\ast}&|v|^2\end{array}\right].
\end{equation}
On can say that von Neumann entropy for the system has Boltzmann one form
\begin{equation}\label{ent}
\mathrm{S}\equiv-\dfrac{\mathrm{tr}(\varrho\ln\varrho)}{\mathrm{tr}(\varrho)}=\ln\left\{\dfrac{1}{2|u|^2-1}\right\}.
\end{equation}
It creates opportunity to formulate of thermodynamics for quantized Einstein--Friedmann Spacetime represented by entropy (\ref{ent}).
\subsection{Thermodynamics of Spacetime}
In opposite to results of paper \cite{g}, in this text we propose use of (\ref{H}) as Hamiltonian of theory. Theory constructed by this Hamiltonian is conceptually simpler than theory presented in \cite{g}, because in (\ref{H}) superfluidity term is absent. However, as it will turn out physical consequences presented below are principally the same as in \cite{g}.

Hamiltonian (\ref{H}) has a following matrix representation in Bogoliubov--Heisenberg basis
\begin{eqnarray}\label{eff}
\mathrm{H}=\left[\begin{array}{cc}\dfrac{|u|^2+|v|^2}{2}\omega&-uv\omega\\-u^{\ast}v^{\ast}\omega&\dfrac{|u|^2+|v|^2}{2}\omega\end{array}\right].
\end{eqnarray}
Statistical mechanics understand internal energy as average $\mathrm{U}=\dfrac{\mathrm{tr}(\varrho\mathrm{H})}{\mathrm{tr}\varrho}$, which for considered case is
\begin{equation}\label{U}
\mathrm{U}=\left(\dfrac{1}{2}+\dfrac{4\mathrm{n}+3}{2\mathrm{n}+1}\mathrm{n}\right)\omega(a).
\end{equation}
Here we introduced quantity
\begin{equation}\label{n}
\mathrm{n}=\langle0|\mathcal{G}^{\dagger}\mathcal{G}|0\rangle=|v(a)|^2=\Bigg|\left|\dfrac{\omega(a_I)}{\omega(a)}\right|-\left|\dfrac{\omega(a)}{\omega(a_I)}\right|\Bigg|^2,
\end{equation}
which is number of particles produced from vacuum. Averaged number of particles is
\begin{equation}
\langle\mathrm{n}\rangle=2\mathrm{n}+1.
\end{equation}
Chemical potential is defined as $\mu=\dfrac{\partial\mathrm{U}}{\partial\mathrm{n}}$, in this case
\begin{equation}\label{mu}
\mu=\left(1+\dfrac{1}{(2\mathrm{n}+1)^2}-\dfrac{1}{2}\dfrac{4\mathrm{n}+1}{4\mathrm{n}^2+2\mathrm{n}}\sqrt{\dfrac{\mathrm{n}}{\mathrm{n}+1}}\right)\omega(a).
\end{equation}
Description of the system according to quantum theory principles demands using of thermal Gibbs ensemble. The system is characterized by CCRs (\ref{ccr1}, \ref{ccr2}). It leads to identification of function under logarithm in (\ref{ent}) with Bose--Einstein partition function
\begin{equation}\label{gib}
\dfrac{1}{2|u|^2-1}\equiv\left\{\exp\left[\dfrac{\mathrm{\mathrm{U}-\mu\mathrm{n}}}{\mathrm{T}}\right]-1\right\}^{-1}.
\end{equation}
Using of relations (\ref{hip}), (\ref{U}), (\ref{n}), (\ref{mu}), and (\ref{gib}) give the formula for temperature of considered system
\begin{eqnarray}
\mathrm{T}=\dfrac{1+\left(\dfrac{2\mathrm{n}}{2\mathrm{n}+1}\right)^2+\dfrac{8\mathrm{n}^2+8\mathrm{n}+1}{4\mathrm{n}+2}\sqrt{\dfrac{\mathrm{n}}{\mathrm{n}+1}}}{2\ln(2\mathrm{n}+2)}\omega(a).
\end{eqnarray}
This formula describes temperature of the system as a function of Friedmann conformal scale factor $a$.
\section{Summary}
Above formalism points at direct way to Quantum Cosmology understood as many-particle open quantum system. On base of this type reasoning formulation of physics is obvious and clear - in presented case description of Einstein--Friedmann Spacetime as quantum Bose--Einstein system leads to well-defined thermodynamics. Furthermore, for considered Spacetime open quantum system point of view gives opportunity to formulate Quantum Gravity as theory with internal structure of logarithmic conformal field theory.

In opinion of the author presented way creates substantially wider opportunities for construction of similar formalism for general gravitational fields and to research connections between geometry and physics in terms of logarithmic conformal field theory.

\end{document}